\documentclass[12pt]{article}

\author{Hao Chen\\
Department of Computing and \\
Information Technology\\
School of Information Science\\
and Engineering\\
Fudan University\\
Shanghai,200433\\
People's Republic of China}
\title{\bf MDS Ideal Secret Sharing Scheme  from \\
AG-codes on Elliptic Curves}
\date{February, 2006}

\begin{document}

\maketitle
\begin{abstract}

For a secret sharing scheme, two parameters $d_{min}$ and
$d_{cheat}$ are defined in [12] and [13]. These two parameters
measure the error-correcting capability and the secret-recovering
capability of the secret sharing scheme against cheaters. Some
general properties of the parameters have been studied in [12],[9]
and [13]. The MDS secret-sharing scheme was defined in [13] and it
was proved that MDS perfect secret sharing scheme can be constructed
for any monotone access structure. The famous Shamir $(k,n)$
threshold secret sharing scheme is the MDS with
$d_{min}=d_{cheat}=n-k+1$. In [3] we proposed the linear secret
sharing scheme from algebraic-geometric codes. In this paper the
linear secret sharing scheme from AG-codes on elliptic curves is
studied and it is shown that many of them are MDS
linear secret sharing scheme.\\

{\bf Index Terms}---  Secret-sharing scheme, MDS secret sharing
scheme, AG-code,  elliptic curve
\end{abstract}

{\bf I. Introduction and Preliminaries}\\

In a secret-sharing scheme among the set of players ${\bf
P}=\{P_1,...,P_n\}$, a dealer $P_0$, not in ${\bf P}$, has a secret,
the dealer distributes the secret among ${\bf P}$ such that only the
qualified subsets of players ${\bf P}$ can reconstruct the secret
from their shares.
 The access structure , $\Gamma \subset 2^{{\bf P}}$, of a secret-sharing scheme is the family of
 the qualified subsets of ${\bf P}$. The minimum accesss structure $min
 \Gamma \subset 2^{{\bf P}}$ is defined to the be the set of minimal elements in
 $\Gamma$(here we use the natural order relation $ S_1 <S_2$ if and
 only if $ S_1 \subset S_2$ on $2^{{\bf P}}$). We call a secret-sharing scheme a $(k,n)$-threshold scheme if the
access
 structure
 consists of the subset of at least  $k$ elements in the set ${\bf P}$, where the number of elements in the set ${\bf P}$
 is
 exactly $n$, that is, among the $n$ players any subset of $k$ or more than $k$ players can reconstruct the secret.
The first secrets-sharing scheme was given independently by Blakley
[2] and Shamir [15] in 1979, actually they gave
 threshold secret-sharing scheme. We call a secret-sharing scheme perfect if the the unqualified
subsets of players to reconstruct the secret have no information of
the secret. The existence of secret-sharing schemes with arbitrary
given access structures was proved in [1] and [8]. Let $K$ be a
finite field, we refer to  [4] for  the definition of linear secret
sharing scheme (LSSS) over $K$ ($K$-LSSS) and its relation with
linear error-correcting
codes. \\

For a secret-sharing scheme, we denote the set of all possible
shares $(v_1,...,v_n)$ (Here $v_i$ is the share of the player $P_i$
for $i=1,...,n$)by ${\bf V}$. Then ${\bf V}$ is a error-correcting
code(not necessarily linear), let $d_{min}$ be the minimum Hamming
distance of this error-correcting code ${\bf V}$. From the
error-correcting capability, it is clear that the cheaters can be
identified from any share(presented by the players)$(v_1,...,v_n)$
if there are at most $[(d_{min}-1]/2]$ cheaters. In [12] McEliece
and Sarwate proved that $d_{min}=n-k+1$ for Shamir's
$(k,n)$-threshold scheme. K.Okada and K.Kurosawa introduced anther
parameter $d_{cheat}$ for general secret-sharing scheme, as the the
number such that the correct secret value $s$ can be recovered if
there are at most $[(d_{cheat}-1)/2]$ cheaters (see [13]). It is
clear that $d_{min} \leq d_{cheat}$. In [13] the authors proved
$d_{cheat}=n-max_{B \in (2^{{\bf P}}-\Gamma})|B|$, where $|B|$ is
the number of the elements in the set $B$. The secret sharing scheme
is called MDS if $d_{min}=d_{cheat}=n-max_{B \in (2^{{\bf
P}}-\Gamma)}$.
It was also proved in [13] that any monotone access structure can be realized by a perfect MDS secret scheme. \\

The approach of secret-sharing based on error-correcting codes was
studied in [4],[5],[9],[10],[11] and [12]. It is found that actually
Shamir's $(k,n)$-threshold scheme is just the secret-sharing scheme
based on the famous Reed-Solomon (RS) code. The error-correcting
code based
 secret-sharing scheme is defined as follow. Here we suppose ${\bf C}$ is
 a linear error-correcting code over the finite
field $GF(q)$ (where $q$ is a prime power) with code length $n+1$
and dimension $k$, i.e., ${\bf C}$ is a $k$ dimension subspace of
$GF(q)^{n+1}$  The Hamming distance $d({\bf C})$ of this
error-correcting code ${\bf C}$ is defined as follows.\\

$$
\begin{array}{ccccccc}
d({\bf C})=min\{wt(v): v \in {\bf C}\}\\
wt(v)=|\{i:v=(v_0,v_1,...,v_n),v_i \neq 0\}|
\end{array}
$$
,where $wt(v)$ is called the Hamming weight of $v$. Let
$G=(g_{ij})_{1 \leq i \leq k ,0 \leq j \leq n}$ be the generator
matrix of ${\bf C}$, i.e., $G$ is a $k \times (n+1)$ matrix in which
$k$ rows of $G$ is a base of the $k$ dimension subspace ${\bf C}$ of
$GF(q)^{n+1}$. Suppose $s$ is a given secret value of the dealer
$P_0$ and the secret is shared among ${\bf P}=\{P_1,...,P_n\}$, the
set of $n$ players .
 Let ${\bf g_1}=(g_{11},...,g_{k1})^T$ be the 1st column of $G$. Chosen a
 random ${\bf u}=(u_1,...,u_k) \in GF(q)^k$ such that $s={\bf u}^{\tau} {\bf g_0}=\Sigma u_ig_{i0}$.
 We have the  codeword ${\bf c}=(c_0,...,c_N)={\bf u}G$, it is clear that $c_0=s$ is the secret, then the dealer $P_0$
 gives the $i-th$ player $P_i$ the $c_i$ as the share of $P_i$ for $i=1,...,n$. In this
 secret-sharing scheme the error-correcting code ${\bf C}$ is assumed to be known to every player and the dealer.
 For a secret sharing scheme from error-correcting codes, suppose that $T_i:GF(q)^k \longrightarrow GF(q)$
  is defined as $T_i({\bf x})= {\bf x}^{\tau} {\bf g_i}$, where $i=0,...,n$ and ${\bf g_i}$ is the $i$-th column of the
  generator matrix of the code ${\bf C}$. In this form we see that the secret sharing scheme is an ideal  linear
  secret sharing scheme over $GF(q)$ ($GF(q)$-LSSS, see [4]). \\

We refer the following Lemma to [5],[10] and [11].\\

 {\em {\bf Lemma 1 }(see [5], [8] and [11]).
  Suppose the dual of ${\bf C}$,
 ${\bf C^{\perp}}=\{v=(v_0,..,v_n): Gv=0\}$ has no codeword of Hamming
 weight 1. In the above secret-sharing scheme based on the error-correcting code
 ${\bf C}$, $(P_{i_1},...,P_{i_m})$ can reconstruct the secret if and only
 if there is a codeword $v=(1,0,...,v_{i_1},...,v_{i_m},...0)$ in ${\bf
 C^{\perp}}$
 such that $v_{i_j} \neq 0$ for at least one $j$, where $1 \leq j \leq m $.}\\

 The secret reconstruction is as follows, since $Gv=0$, ${\bf
 g_1}=-\Sigma_{j=1}^m v_{i_j} {\bf g_{i_j}}$, where ${\bf g_h}$ is
 the $h-th$ column of $G$ for $h=1,...,N$. Then $s=c_0={\bf u}{\bf
 g_1}=-{\bf u}\Sigma_{j=1}^m  {\bf g_{i_j}}=-\Sigma_{j=1}^m
 v_{i_j}c_{i_j}$.\\

We need recall some basic facts about algebraic-geometric codes. Let
${\bf X}$ be an absolutely irreducible, projective and smooth curve
defined over $GF(q)$ with genus $g$, ${\bf D}=\{P_0,...P_n\}$ be a
set of $GF(q)$-rational points of ${\bf X}$ and ${\bf G}$ be a
$GF(q)$-rational divisor satisfying $supp({\bf G})\bigcap {\bf
D}=\emptyset$. Let $L(G)=\{f: (f)+G \geq 0 \}$ is the linear space
(over $GF(q)$) of all rational functions with its divisor not
smaller than $-G$ and $\Omega(B)=\{\omega: (\omega) \geq B\}$ be the
linear space of all differentials with their divisors not smaller
than $B$. Then the functional AG(algebraic-geometric )code ${\bf
C_L(D,G)} \in GF(q)^{n+1}$ and residual AG(algebraic-geometric) code
${\bf C_{\Omega}(D,G)} \in GF(q)^{n+1}$ are defined. ${\bf
C_L(D,G)}$ is a $[n+1,k=dim(L({\bf G})-dim(L({\bf G}-{\bf D}), d
\geq n+1-deg({\bf G})]$ code over $GF(q)$ and ${\bf
C_{\Omega}(D,G)}$ is a $[n+1,k=dim(\Omega({\bf G}-{\bf
D}))-dim(\Omega({\bf G})),d \geq deg({\bf G})-2g+2]$ code over
$GF(q)$. We know that the functional code is just the evaluations of
functions in $L(G)$ at the set ${\bf D}$ and the residual code is
just the residues of
differentials in $\Omega(G_D)$ at the set ${\bf D}$ (see [16], [17] and 18]).\\

We also know that ${\bf C_L(D,G)}$ and ${\bf C_{\Omega}(D,G)}$ are
dual codes. It is known that for a differential $\eta$ that has
poles at $P_1,...P_n$ with residue 1 (there always exists such a
$\eta$, see[16]) we have ${\bf C_{\Omega}(D,G)}={\bf
C_L(D,D-G+(\eta))}$, the function $f$ corresponds to the
differential $f\eta$. This means that  functional codes and residue
code are essentially same. \\

{\bf II. Main Results}\\

Let ${\bf X}$ be an absolutely irreducible, projective and smooth
curve defined over $GF(q)$ with genus $g$, ${\bf D}=\{P_0,...P_n\}$
be a set of $GF(q)$-rational points of ${\bf X}$ and ${\bf G}$ be a
$GF(q)$-rational divisor with degree $m$ satisfying $ supp({\bf G})
\bigcap {\bf D}=\emptyset$. We can have a LSSS on the $n$ players
${\bf P}=\{P_1,...,P_n\}$ from the linear code ${\bf
C_{\Omega}(D,G)}$, thus we know that the reconstruction of the
secret is based from its dual code ${\bf C_L(D,G)}$. For the curve
of genus 0 over $GF(q)$, we have exactly the same LSSS as Shamir's
$(k,n)$-threshold scheme, since the AG-codes over the curve of genus
0 is just the RS codes
(see [16],[17] and 18]).\\

The following  Theorem 4 and Corollary 1 are the main results of this paper.\\

{\em {\bf Theorem 1.} For the LSSS over $GF(q)$ from the code ${\bf
C_{\Omega}(D,G)}$ we have $m-2g+1 \leq d_{min} \leq d_{cheat} \leq
m+1$.}\\

{\bf Proof.} From the theory of AG-codes ([12-14]), we know ${\bf
C_{\Omega}(D,G)}$ can be identified with ${\bf C_L(D,D-G+(\eta))}$.
Thus $d_{min}$ is the minimum Hamming weight of ${\bf C_L(P,D-G+(\eta))}$. We have $d_{min} \geq m-2g+1$.\\

On the other hand any subset of ${\bf P}$ less than $n-m$ elements
is not qualified from the fact that the minimum Hamming weight of
${\bf C_L(D,G)}$ is $n+1-m$. From the equality $d_{cheat}=n-max_{B
\in 2^{{\bf P}}-\Gamma} |B|$, we have $d_{cheat} \leq
n-(n-m-1)=m+1$. The conclusion is proved.\\

We need to recall the following result in [14].\\

{\em {\bf Theorem 2 (see [14] and [7]).} 1).Let $E$ be an elliptic
curve over $GF(q)$ with the group of $GF(q)$-rational points
$E(GF(q))$. Then $E(GF(q))$ is isomorphic to $Z_{n_1} \bigoplus
Z_{n_2}$, where $n_1$
is a divisor of $q-1$ and $n_2$\\
2) If $E$ is supersingular, then $E(GF(q))$ is either\\
a)cyclic; \\
b)or $Z_{2} \bigoplus Z_{\frac{q+1}{2}}$; \\
c)or $Z_{\sqrt q-1} \bigoplus Z_{\sqrt q-1}$;\\
d)or $Z_{\sqrt q+1} \bigoplus Z_{\sqrt q+1}$.}\\

For any given elliptic curve $E$ over $GF(q)$, let
$D'=\{g_0,g_1,...g_{H}\}$ be a subset of $E(GF(q))$ of $H+1$
non-zero elements, let $G=mO$ ($O$ is the point of the zero element
of $E(GF(q))$). $g_0,...,g_{H}$ correspond to the rational points
$P_0,P_1,...,P_{H}$ of $E(GF(q))$. In the construction, we take
${\bf D}=D'$ and ${\bf
P}=\{P_1,...,P_{H}\}$. We have the following result.\\

{\em {\bf Theorem 3.} a) Let $A=\{P_{i_1},...,P_{i_t}\}$ be a subset
of ${\bf P}$ with $t$ elements, $B$ is the element in $E(GF(q))$
such that the group sum of $B$ and $g_{i_1},...,g_{i_t}$ is zero in
the group $E(GF(q))$. Then $A^c$ (Here $A^c$ is the set ${\bf P}-A$
) is a qualified subset for
the LSSS from ${\bf C_{\Omega}(D,G)}$ only if $t \leq m$ and  \\
1) When $t=m$, $A^c$ is a minimal qualified subset if and only if $B=O$, the zero element of $E(GF(q))$;\\
2) When $t=m-1$, $A^c$ is a minimal qualified subset  if and only if $B$ is not in ${\bf D}$ or $B$ is in the set $A$.\\
b) Any subset of ${\bf P}$ of more than $n-m+2$ elements is qualified.}\\

{\bf Proof.} From the theory of AG-codes, the minimum Hamming weight
of ${\bf C_L(D,G)}$ is $n+1-m$, thus $A^c$ is a qualified subset
only if $t \leq m$.\\

We know that for any $t$ points $W_1,...,W_t$ in $E(GF(q))$ the
divisor $W_1+...+W_t- tO$ is linear equivalent to the divisor $W-O$,
where $W$ is the group sum of $W_1,...,W_t$ in the group $E(GF(q))$.
$\{P_{i_1},...,P_{i_m}\}^c$ is a qualified subset (therefor minimal
qualified subset) if there exist a function $f \in L(G)$ such that
$f(P_{i_1})=...=f(P_{i_m})=0$, this means that the divisor
$P_{i_1}+...+P_{i_m}$ is linearly equivalent to ${\bf G}$.
The conclusion of a) is proved. \\

$\{P_{i_1},...,P_{i_{m-1}}\}^c$ is a qualified subset if there exist
a function $f \in L(G)$ such that $f(P_{i_1})=...=f(P_{i_{m-1}})=0$,
this means that the divisor $P_{i_1}+...+P_{i_{m-1}}+B'$ is linearly
equivalent to ${\bf G}$ for some effective divisor $B'$.  It is
clear that $deg(B')=1$ and $B'$ is a $GF(q)$-rational point in $E$.
Thus $B'$ is just the $B$ in the condition.  On the other hand we
note that $B \neq P_0$, so $B$ has to be in $A$ or a point not in
${\bf D}$. The
conclusion of a) is proved.\\

If $A$ is a subset of ${\bf P}$ such that $|A| \leq m-2$, the
divisor ${\bf G}-A$ has its degree $deg({\bf G}-A) \geq 2$. So the
corresponding system has no base point. We can find a function in
$L({\bf G}-A)$ such that it is not zero at $P_0$, thus we have a
codeword in ${\bf C_L(D,G)}$ which is not zero at $P_0$ and zero at
all points of $A$. This implies that $A^c$ is a qualified subset. The
conclusion of b) is proved.\\

The following Corollary is a direct result of Theorem 3.\\

{\em {\bf Corollary 1.} If there is a subset of ${\bf P}$ of $H-m+1$
elements which is not $A^c$ of type 2) as in the above Theorem 3 and
do not contain any subset of $H-m$ elements of type a) in Theorem 3,
then the LSSS in Theorem 1 is MDS (perfect) ideal
secret sharing scheme.}\\

{\em {\bf Theorem 4.} If ${\bf D}\bigcup \{O\}$ is a subgroup of $E(GF(q))$,
then the ideal LSSS in Theorem 3 is MDS.}\\

{\bf Proof.} We prove that there exist $m-1$ distinct elements
$g_{i_1},...,g_{i_{m-1}}$ in ${\bf P}$ such that
$g_{i_1}+...+g_{i_{m-1}}=-g_0$. First we choose 2 elements
$g_{i_1},g_{i_2}$ when $m-1$ is even ( or 3 elements
$g_{i_1},g_{i_2},g_{i_3}$ when $m-1$ is odd) in the group ${\bf D}
\bigcup \{O\}$ such that $g_{i_1}+g_{i_2}=-g_0$
($g_{i_1}+g_{i_2}+g_{i_3}=-g_0$ when $m-1$ is odd). The other $m-3$
(when $m-1$ is even, or $m-4$ when $m-1$ is odd) elements can be
taken to be pairs of elements $(g_{i_j},-g_{i_j})$. Since
${\bf D} \bigcup \{O\}$ is group, thus the desired points can always be found.\\

For this subset $A$ of $m-1$ elements in ${\bf P}$, if it is
qualified we know that $B$ in Theorem 3 is $P_0$, this is a
contradiction to Theorem 3. We have a
subset of ${\bf P}$ of $n-m+1$ elements which is not qualified. This implies $d_{cheat} \leq m-1$.
From Theorem 1 $m-1 \leq d_{min} \leq d_{cheat} \leq m-1$, we have $d_{min}=d_{cheat}=m-1$. The conclusion is proved.\\

{\bf III. Examples}\\

{\bf Example 1. }Let $E$ be the elliptic curve $y^2=x^3+5x+4$
defined over $GF(7)$. Then $E(GF(7))$ is a cyclic group of order
$10$ with $O$ the point at infinity and
$P_0=(3,2),P_1=(2,6),P_2=(4,2),P_3=(0,5)$
$P_4=(5,0),P_5=(0,2),P_6=(4,5),P_7=(2,1),P_8=(3,5)$. From an easy
computation we know that $P_0$ is a generator of $E(GF(7))$ and
$P_i$ is $(i+1)P_0$ (in the group operation of $E(GF(7))$.) We take
${\bf G}=3O,{\bf D}=\{P_0,P_1,P_3,P_5,P_7\}$, then the access
structure of the ideal $GF(7)$-LSSS from ${\bf
C_{\Omega}(D,G)}$ are the following subsets of ${\bf P}=\{P_1,P_3,P_5,P_7\}$.\\
1) All subsets of ${\bf P}$ with $3$ elements and the set ${\bf P}$;\\
2) The following $6$ subsets of $2$ elements $\{P_1,P_7\}$,
$\{P_1,P_3\}$, $\{P_1,P_5\}$,
$\{P_3,P_5\}$, $\{P_3,P_7\}$, $\{P_5,P_7\}$ are minimal qualified subsets.\\

We can check that every subset of ${\bf P}$ of $2$ elements is
qualified so $d_{cheat}=3$, it is easy to see that $d_{min}=2$ we
conclude that this ideal LSSS is not MDS.\\

{\bf Example 2.} Let $E$ be the elliptic curve $y^2+y=x^3$ defined
over $GF(4)$. This is the Hermitian curve over $GF(4)$, it has $9$
rational points and $E(GF(4))$ is isomorphic to $Z_3 \bigoplus Z_3$.
We take ${\bf G}=3O$, where $O$ is the zero element in the group
$E(GF(4))$. Let $P_{ij}$ be the rational point on $E$ corresponding
to $(i,j)$ in $Z_3 \bigoplus Z_3$. ${\bf
D}=\{P_{10},P_{01},...,P_{22}\}, {\bf P}=\{P_{01},...,P_{22}\}$.\\
Then the qualified subsets of ${\bf P}$ are as follows.\\
1) The qualified subsets of $4$ elements are
$\{P_{20},P_{21},P_{02}\}^c$, $\{P_{01},P_{20},P_{22}\}^c$, $\{P_{11},P_{12},P_{20}\}^c$.\\
2) The qualified subsets of $5$ elements are
$\{P_{01},P_{02}\}^c$, $\{P_{11},P_{22}\}^c$, $\{P_{12},P_{21}\}^c$.\\
3) The subsets of ${\bf P}$ of $6$ elements and the set ${\bf P}$
are qualified.\\

The subsets in 1) and 2) are the minimal qualified subsets. It is clear that
$d_{min}=m-2g+1=2$ and $d_{cheat}=7-5=2$. Thus this ideal LSSS is MDS.\\

{\bf Example 3.} Let $E$ be the elliptic curve $y^2+y=x^3$ defined
over $GF(q),q=2^r$. This is a super-singular  elliptic curve,
$E(GF(q))$ has $2^r+1$ rational points and is isomorphic to a cyclic
group when $r$ is an odd number; $E(GF(q))$ has $2^r+1+2 \cdot
2^{\frac{r}{2}}$ rational points and is isomorphic to the product of
two cyclic groups of order $2^{\frac{r}{2}}+1$ when $r$ is an even
number.  We take ${\bf G}=mO$, where $O$ is the zero element in the
group $E(GF(q))$. Let ${\bf D}$ be the set of all non-zero rational
points and the point $P_0$ be an arbitrary non-zero point in ${\bf
D}$.
From Theorem 4, the ideal LSSS over $GF(q)$ is MDS.\\

For any fixed $r$, we can calculate the access structure as in
Example 2. Now suppose $r=3$. Then the access structure can be
computed as follows.\\

In the case over $GF(8)$, $E(GF(8))$ has $9$ rational points and it
is a cyclic group of order $9$. Let $P_{i}$ be the rational point on
$E$ corresponding to $i$ in $Z_9=\{0,1,2,...,7,8\}$ for
$i=1,2...,8$. Let ${\bf G}=3O$, where $O$ corresponds to the zero
element $0$ in the group $E(GF(8))$,  ${\bf D}=\{P_{1},...,P_{8}\}$
and ${\bf P}=\{P_2,...,P_8\}$. Then the access structure of the
ideal LSSS
from ${\bf C_{\Omega}(D,G)}$ is as follows.\\
1) The minimal qualified subsets of $4$ elements are
$\{P_{2},P_{3},P_{4}\}^c$,$\{P_{3},P_{7},P_{8}\}^c$, $\{P_4,P_6,P_8\}^c$, $\{P_{5},P_{6},P_{7}\}^c$.\\
2) The minimal qualified subsets of $5$ elements are
$\{P_{2},P_{5}\}^c$,$\{P_{2},P_{7}\}^c$, $\{P_{2},P_{8}\}^c$,
$\{P_{3},P_{6}\}^c$,
$\{P_{4},P_{5}\}^c$, $\{P_{4},P_{7}\}^c$, $\{P_5,P_8\}^c$.\\
3) The subsets of ${\bf P}$ of $6$ elements and the set ${\bf P}$
are qualified.\\

{\bf IV. Conclusion}\\

We have proved some sufficient conditions about the MDS ideal linear
secret-sharing scheme from the AG-codes on elliptic curves, which
can be thought as a natural generalization of Shamir's
$(k,n)$-threshold scheme(from AG-codes on the genus 0 curve, RS
codes). From the main results of this paper many MDS ideal secret
sharing schemes can be constructed. This demonstrates that elliptic
curves, perhaps also hyper-elliptic curves, are important resource
in the theory and practice of
secret-sharing.    \\

{\bf Acknowledgement.} This work was supported in part by  NNSF,
China under Grant 90607005 and Distinguished Young Scholar Grant
10225106.\\

\begin{center}
REFERENCES
\end{center}

[1] J.Benaloh and J.Leichter, Generalized secret sharing and
monotone functions, Crypto'88, LNCS-403, pp.25-35.\\

[2] G.R.Blakle, Safeguarding cryptographic keys, Proc. NCC
AFIPS, pp.313-317, 1979.\\

[3] Hao Chen, Linear secret sharing from algebraic-geometric codes,
preprint 2005, submitted to Crypto 2006, accepted and merged with R.Cramer's paper, see LNCS 4117.\\

[4] R. Cramer, V.Daza,I.Cracia, J.J. Urroz, C.Leander, J.Marti-Farre
and C.Padro, On Codes, matroids and secure multi-party computations
from linear secret sharing schems, Advances in Cryptology, Crpto
2005, LNCS 3621, pp327-343.\\

[5] C.Ding, D.R.Kohel and S.Ling, Secret-sharing with a class of
tenary codes, Theoretical Computer Science, vol.246,
pp.285-298, 2000.\\

[6] R.Hartshorne, Algebraic geometry GTM 52, Springer-Verlag, 1977.\\

[7] D.Hankerson, A.Menezes and S.Vanstone, Guide to elliptic curve
cryptography, Springer-Verlag, 2004.\\

[8] M.Ito, A.Saito and T.Nishizeki, Secret sharing scheme realizing
general access structures, Proc. IEEE Globalcom'87, Tokyo,
pp.99-102\\

[9] E.D.Karnin, J.W.Green and M.E.Hellman, On secret sharing
systems, IEEE Transactions on Information Theory,vol.29, no.1,
pp.35-41, Jan. 1982.\\

[10] J.L.Massey, Minimal codewords and secret sharing, Proc. 6th
Joint Sweidish-Russsian workshop on Information Theory, Molle,
Sweden,August 22-27, 1993,pp269-279. \\

[11] J.L.Massey, Some applications of coding theory in cryptography,
in P.G.Farrell (Ed.0, Codes and Ciphers: Cryptography and Coding IV,
Formara Ltd, Essses, England, 1995, pp.33-47.\\

[12] R.J.McEliece and D.V.Sarwate, On sharing secrets and Reed-Solomom codes, Comm. ACM,
22,11, pp.612-613, Nov.1979.\\

[13] K.Okada and K.Kurosawa, MDS secret-sharing scheme secure
against cheaters, IEEE Transactions on Information Theory,
vol.46, no.3, pp.1078-81, April 2000.\\

[14] R.Schoof, Nonsingular plane cubic curves over finite fields,
J.Combin. Theory, A, vol.46, pp.183-211, 1987.\\

[15] A.Shamir, How to share a secret, Comm. ACM 22(1979), pp.612-613\\

[16] H.Stichtenoth, Algebraic function fields and codes, Springer,
Berlin, 1993.\\

[17] M.A.Tsfasman and S.G.Vladut, Algebraic-geometric codes, Kluwer,
Dordrecht, 1991.\\

[18] J.H.van Lint, Introduction to coding theory (3rd Edition),
Springer-Verlag, 1999.\\

\end{document}